\documentclass[aps,prd,twocolumn,reprint,preprintnumbers,floatfix,nofootinbib,superscriptaddress]{revtex4-1}

\input{header}
\usepackage{ulem}
\usepackage{float}

\def\carleton{Department of Physics, Carleton University, Ottawa, ON K1S 5B6, Canada }

\newcommand{\LmuLtau}{U(1)_{L_\mu - L_\tau}}

\begin{document}
\date{\today}
\title{$B$ Anomalies and Dark Matter in an $L_\mu - L_\tau$ Model with General Kinetic Mixing} 

\author{Douglas Tuckler}
\email{dtuckler@physics.carleton.ca}
\affiliation{\carleton}

\begin{abstract}
We revisit the $L_\mu - L_\tau$ extension of the Standard Model that can simultaneously address anomalies in semileptonic $B$ meson decays and the nature of dark matter (DM). In the region favored by the $B$ anomalies, this scenario is excluded by a combination of low-energy flavor constraints and stringent DM direct detection constraints if the kinetic mixing between $Z^\prime$  gauge boson and the photon vanishes at high energy scales, since this leads to a sizable coupling between DM and the SM in low-momentum scattering processes. However, this is no longer the case if the kinetic mixing vanishes at low energy scales instead. In this scenario,  the low-momentum DM scattering rate can be suppressed and constraints from direct detection experiments can be substantially relaxed. As a result, we find a re-opening  of the region of parameter space where the $B$ anomalies and DM can be simultaneously explained. 

\end{abstract}

\maketitle

\section{Introduction}

The $Z^\prime$ gauge boson of a broken $U(1)_{L_\mu - L_\tau}$ symmetry is a simple extension of the Standard Model (SM) \cite{He:1991qd,He:1990pn} that can address the tension in the anomalous magnetic moment of the muon $(g-2)_\mu$ \cite{Baek:2001kca}, can serve as mediator for dark matter (DM) interactions with the SM \cite{Altmannshofer:2016jzy,Arcadi:2018tly,Kamada:2018zxi,Foldenauer:2018zrz,Borah:2021jzu,Holst:2021lzm}, and can address anomalies observed in the decays of $B$ mesons \cite{Altmannshofer:2014cfa, Crivellin:2015mga,Altmannshofer:2016jzy,Crivellin:2016ejn,Ko:2017yrd,Arcadi:2018tly}. In fact, a virtue of the $U(1)_{L_\mu - L_\tau}$ extension of the SM is that in can address more than one of these questions at the same time.

Previously, it was shown that the $Z^\prime$ can address both the $B$ anomalies and DM \cite{Altmannshofer:2016jzy}. By introducing a flavor-violating coupling of the $Z^\prime$ to left-handed $b$ and $s$ quarks the $Z^\prime$ is able to modify the $b \to s \mu \mu$ transition as preferred by global fits to rare $B$ decays  \cite{Altmannshofer:2014cfa,Altmannshofer:2015mqa,Altmannshofer:2021qrr}, and the kinetic mixing between the $Z^\prime$ and the SM photon allows for $L_\mu - L_\tau$ charged DM to interact with SM particles. Strong constraints from DM direct detection experiments allowed for only a narrow window where the DM annihilates nearly on resonance in the early Universe \cite{Altmannshofer:2016jzy}.  As we will see, the latest DM direct detection results completely exclude this scenario.

However, the interpretation of these direct detection constraints depends strongly on the energy dependence of the kinetic mixing. Typically, it is assumed that the kinetic mixing vanishes at energy scales much higher than the $\tau$ lepton mass, which leads to a sizable coupling at low energies between DM and nuclei of direct detection experiments. This assumption is by no means general and other boundary conditions are equally viable.

Recently, it was shown that DM direct detection constraints can be substantially weakened if the kinetic mixing is suppressed at energy scales lower than the muon mass \cite{Hapitas:2021ilr}. It was shown in \cite{Hapitas:2021ilr} that this scenario can open up new parameter space for a simultaneous explanation of DM and the $(g-2)_\mu$ anomaly. 

 In this paper, we revisit the explanation of the $B$ anomalies and DM in  the $\LmuLtau$ model by considering this alternative treatment of the kinetic mixing between the $Z^\prime$ and the SM photon. We will show that stringent constrains from DM direct detection experiments can be relaxed if the kinetic mixing is suppressed at low momentum transfer. We find a reopening of viable parameter space that addresses the $B$ anomalies and DM simultaneously, consistent with all other existing constraints from neutrino trident production, collider searches, and DM indirect detection.

This paper is organized as follows. In Sec.~\ref{sec:model} we introduce the  $\LmuLtau$ model and discuss the momentum dependence of the kinetic mixing between $Z^\prime$ and the photon. In Sec.~\ref{sec:Banomalies} we discuss the connection between the model under consideration and the $B$ anomalies, as well as flavor constraints that arise from the flavor-violating quark coupling. Sec.~\ref{sec:DM} focuses on the DM phenomenology, updating the model of \cite{Altmannshofer:2016jzy} with the latest DM direct detection constraints and showing the effect of the momentum dependence of the kinetic mixing on the these constraints. In Sec.~\ref{sec:EWPO} we compute the constraints from shifts in the $Z$ boson mass induced by the kinetic mixing with the $Z^\prime$ at high scales. We conclude in Sec.~\ref{sec:conclusion}.

\section{The $L_\mu - L_\tau$ Model}\label{sec:model}

Following \cite{Altmannshofer:2016jzy}, we consider the gauged $U(1)_{L_\mu - L_\tau}$ extension of the SM, whose symmetry breaking gives rise to a $Z^\prime$ gauge boson with mass $m_{Z^\prime}$.\footnote{We remain agnostic about the details of this symmetry breaking and assume that the physics involved is sufficiently decoupled, having no effect on the phenomenology discussed in this paper.} In addition, we introduce a Dirac fermion DM candidate $\chi$ with mass $m_\chi$ and charged under the $U(1)_{L_\mu - L_\tau}$ symmetry. We also introduce a  flavor-violating interaction of the $Z^\prime$ to left-handed $b$ and $s$ quarks. The Lagrangian describing this model is
\begin{equation}\label{eq:Lag}
\begin{split}
\mathcal{L}&_{L_\mu - L_\tau} \supset  \ - \frac{1}{4} Z^\prime_{\alpha\beta} Z'^{\alpha\beta}  + \frac{1}{2} m^2_{Z^\prime} Z^\prime_\alpha Z^{\prime \alpha} + \frac{\varepsilon_0}{2} Z^\prime_{\alpha\beta}  F^{\alpha\beta}  \\
 & + q_\ell g^\prime\big( \bar\mu \gamma^\alpha \mu - \bar\tau \gamma^\alpha \tau  
+  \bar\nu_\mu \gamma^\alpha P_L\nu_\mu -  \bar\nu_\tau \gamma^\alpha P_L \nu_\tau\big) Z'_\alpha  \  \\
& + i\bar \chi  \gamma^\alpha \partial_\alpha \chi - m_\chi \bar \chi \chi + q_\chi g^\prime \bar\chi \gamma^\alpha \chi Z^\prime_\alpha  \ \\
& + g_{bs}\bar{s} \gamma^\alpha P_L b Z^\prime_\alpha \ ,
\end{split}
\end{equation}
where $g^\prime$ is the $\LmuLtau$ coupling constant, $q_\ell$ and $q_\chi$ are the charge of the leptons and DM, respectively, under $\LmuLtau$, and $g_{bs}$ is the flavor-violating coupling of $Z^\prime-b-s$ interaction. Note that $q_{\ell,\chi}$ are free parameters but we will simply assume $q_\chi = q_\ell = 1$. 

In Eq.~(\ref{eq:Lag}) we have introduced a bare kinetic mixing parameter $\varepsilon_0$ between $Z^\prime$ and the photon field strengths. This is a free parameter of the model and depends on unknown ultraviolet (UV) physics. However, we will see that it can be fixed by considering the asymptotic boundary conditions of the total kinetic mixing in, for example, DM-nucleon scattering.

\subsection{Momentum Dependence in Kinetic Mixing}
\begin{figure}[t]
\centering
\includegraphics[width=0.45\textwidth]{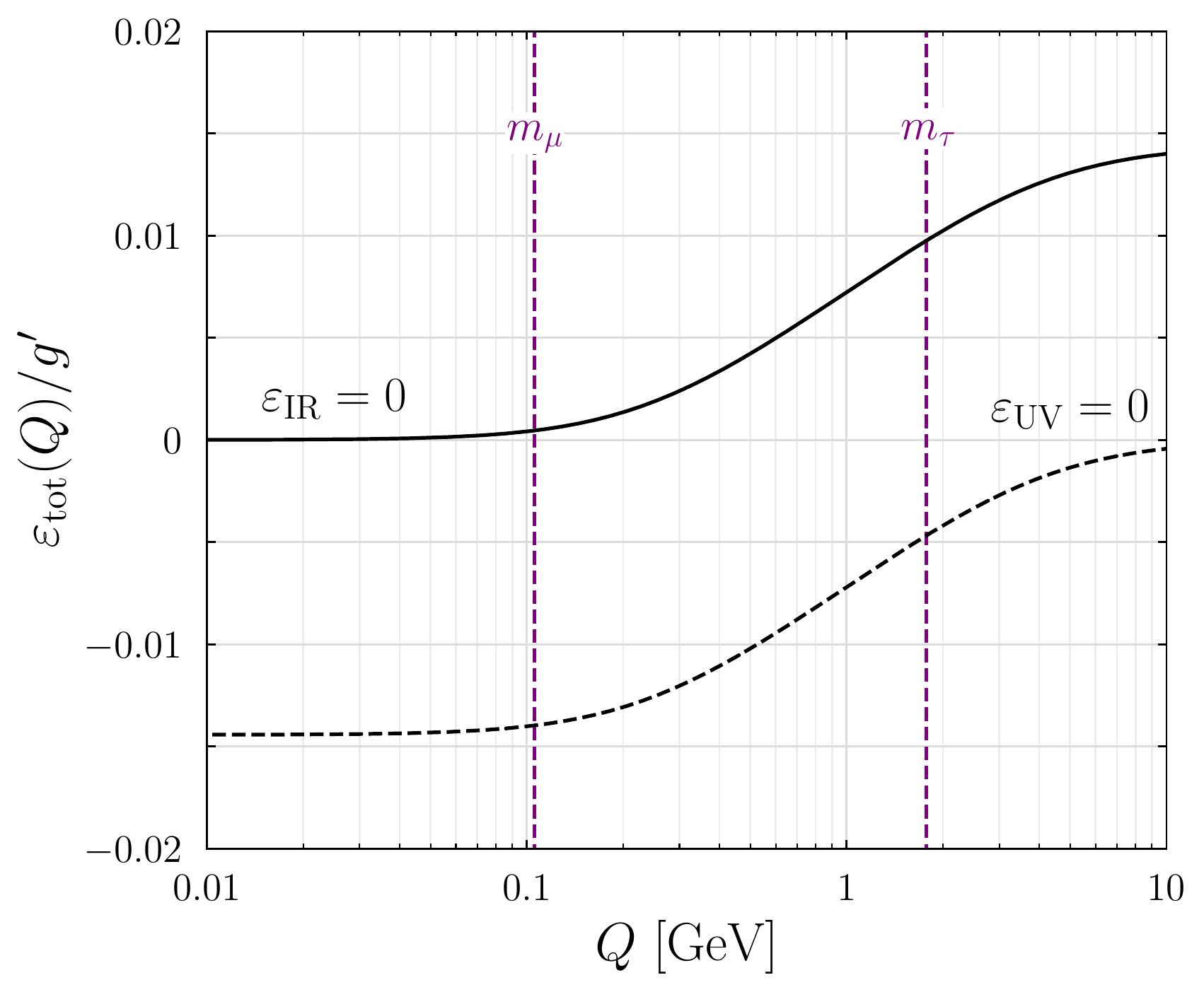}
\caption{Momentum dependent kinetic mixing $\varepsilon_\text{tot}(Q)$ as a function of momentum transfer $Q$. The solid (dashed) curve corresponds to the boundary condition $\varepsilon_\text{IR} = 0$ of Eq.~(\ref{eq:IR}) ($\varepsilon_\text{UV} = 0$ of Eq.~(\ref{eq:UV})). The vertical, dashed purple lines indicate where the momentum transfer is equal to $m_\mu$ and $m_\tau$. }\label{fig:running}
\end{figure}

In this model, the $Z^\prime$ gauge boson does not couple directly to electrons and quarks of the SM, and the scattering of DM (and neutrinos) with nucleons of a detector is induced by the kinetic mixing between the $Z^\prime$ and the photon. The total kinetic mixing has a tree-level contribution $\varepsilon_0$ and a one-loop contribution from virtual $\mu$ and $\tau$ leptons. The total kinetic mixing with explicit momentum transfer dependence is given by
\begin{equation}\label{eq:epsilontotal}
\varepsilon_{\rm tot} (Q) = \varepsilon_0  - \frac{e g^\prime}{2\pi^2} \int_0^1 dx \, x(1-x) \log \left[ \frac{m_\tau^2 + x(1-x)Q^2}{m_\mu^2 + x(1-x)Q^2} \right] \ , 
\end{equation}
where $Q^2 = \sqrt{(-q^2)} > 0$, with $q^\mu$ being the momentum transfer in the scattering process. The bare kinetic mixing parameter $\varepsilon_0$ can be fixed by considering boundary conditions of the total kinetic mixing. Here we consider two boundary conditions:
\begin{align}
\varepsilon_{\rm tot} (Q) &\to 0~\text{as}~Q \to \infty ~~\Rightarrow~~\varepsilon_0 = 0 \label{eq:UV}\\
\varepsilon_{\rm tot} (Q) &\to 0~\text{as}~Q \to 0 ~~~\Rightarrow~~\varepsilon_0 = \frac{eg^\prime}{12\pi^2}\log\frac{m^2_\tau} {m^2_\mu}  \label{eq:IR} \ ,
\end{align}
which fixes the bare kinetic mixing parameter $\varepsilon_0$ to the values after the double right arrows in Eqs.~(\ref{eq:UV}) and (\ref{eq:IR}). For brevity, we will denote the boundary condition in Eq.~(\ref{eq:UV}) by $\varepsilon_\text{UV} = 0$ (i.e $\varepsilon_{\rm tot}$ vanishes at high or UV scales), and the boundary condition in Eq.~(\ref{eq:IR}) by $\varepsilon_\text{IR} = 0$ (i.e $\varepsilon_{\rm tot}$ vanishes at low or  infrared (IR) scales).

 The boundary condition in Eq.~(\ref{eq:UV}) is one that is typically assumed in the literature, but it is by no means a generic choice. Recently, it was shown that varying the boundary condition of the total kinetic mixing has non-trivial effects the phenomenology of neutrino and DM scattering in the $L_\mu - L_\tau$ model \cite{Hapitas:2021ilr}. In particular, the boundary condition in Eq.~(\ref{eq:IR}) results in a suppression of the total kinetic mixing  at low momentum transfer, as depicted in Fig.~(\ref{fig:running}).

 An important consequence of this behavior is that constraints from processes that occur at low momentum transfer, such as those from coherent elastic neutrino-nucleus scattering (CE$\nu$NS) and DM direct detection experiments, can be substantially weaker and regions of parameter space that would otherwise be severely constrained become viable. We will explore this for DM direct detection in Sec.~\ref{sec:DM}.


\section{$B$ Anomalies and Constraints}\label{sec:Banomalies}

In this section we discuss the how the $Z^\prime$ of the $\LmuLtau$ model can address the $B$ anomalies. We also briefly discuss constraints on the parameter space from meson mixing, neutrino trident production, and $Z \to 4\mu$ searches.

\subsection{$B$ Anomalies}
Over the past decade, $B$ factories such as Belle, LHCb, ATLAS, and CMS have observed anomalies in rare semileptonic decays of $B$ mesons. Discrepencies between experimental results and SM predictions have been seen in differential branching ratios, angular observables, and lepton flavor universality (LFU) ratios of various $b\to s$ transitions such as $B^0 \to K^{\ast 0} \ell^+ \ell^-$,  $B^+ \to K^+ \ell^+ \ell^-$, and  $B_s \to \phi \ell^+ \ell^-$ \cite{Belle:2009zue,CDF:2011buy,BaBar:2012mrf,LHCb:2014cxe,LHCb:2015tgy,LHCb:2015wdu,CMS:2015bcy,LHCb:2016ykl,Belle:2016fev,LHCb:2017avl,CMS:2017rzx,ATLAS:2018gqc,LHCb:2019hip,Belle:2019oag,BELLE:2019xld,LHCb:2020lmf,LHCb:2020gog,LHCb:2021trn,LHCb:2021zwz,LHCb:2021lvy}.

Global fits to all rare $B$ decay data find a strong preference for new physics in the form a four-fermion contact interaction described by the effective Hamiltonian
\begin{equation}\label{eq:Heff}
\mathcal{H}^\text{NP}_\text{eff} = -\frac{4 G_F}{\sqrt{2}}V_{tb} V^\ast_{ts} \frac{e^2}{16\pi^2} C^\text{NP}_9(\bar{s} \gamma_\alpha P_L b)(\bar{\mu} \gamma^\alpha \mu)
\end{equation}
with a best fit value for the Wilson coefficient of $C_9^\text{NP} = -0.73$ \cite{Altmannshofer:2021qrr}. This kind of interaction can be accommodated in the $\LmuLtau$ model since the $Z^\prime$ has the required vector coupling to muons, but does not couple to electrons to leading order. Integrating out the $Z^\prime$ in Eq.~(\ref{eq:Lag}) leads to the contact interaction in Eq.~(\ref{eq:Heff}) and the Wilson coefficient is determined by $g^\prime$, $m_{Z^\prime}$, and $g_{bs}$ as
\begin{equation}
C^\text{NP}_9 = - \frac{g^\prime g_{bs}}{2 m^2_{Z^\prime}} \frac{\sqrt{2}}{4 G_F} \frac{16 \pi^2}{e^2}\frac{1}{V_{tb} V^\ast_{ts}} \ .
\end{equation}

We will fix the value of $g_{bs}$ such that this always holds i.e. we have
\begin{equation}
g_{bs} = - \frac{4 G_F}{\sqrt{2}}V_{tb} V^\ast_{ts} \frac{e^2}{16\pi^2} C^\text{NP}_9 \frac{2 m^2_{Z^\prime}}{g^\prime} \simeq \frac{10^{-9}}{\text{GeV}^2} \times \frac{m^2_{Z^\prime}}{g^\prime}\ ,
\end{equation}
so that the phenomenology of this model is entirely determined by $g^\prime$ and $m_{Z^\prime}$. To successfully address the $B$ anomalies $m_{Z^\prime}$ cannot be too small. If $m_{Z^\prime}$ is smaller than the $B$ meson mass, it could be produced on-shell in $B$ decays and would appear as a resonance in the dimuon invariant mass distribution. The non-observation of such resonance by LHCb \cite{LHCb:2015nkv} restricts the $Z^\prime$ mass to be above the $B$ meson mass, and we will conservatively require $m_{Z^\prime} \gtrsim 10$ GeV to explain the $B$ anomalies. In Fig.~(\ref{fig:DD}) this bound is denoted by the vertical dashed, black line at $m_{Z^\prime} = 10$ GeV. Note, the choice of  $m_{Z^\prime} \gtrsim 10$ GeV also allows us to use the EFT framework described in Eq.~(\ref{eq:Heff}).
\subsection{Flavor Constraints}
In addition to the $b \to s \mu\mu$ transition discussed above, the flavor violating coupling in the last line of Eq.~(\ref{eq:Lag}) will contribute at tree-level to low energy flavor observables. In particular, it will lead to a tree-level contribution to the $B_s$ mass difference $\Delta M_s$. The mass difference is parameterized by the mixing amplitude $M_{12}$ as
\begin{equation}
\Delta M_s = \Delta M^\text{SM}_s  \Bigg| 1 + \frac{M^{Z^\prime}_{12}}{M^\text{SM}_{12}} \Bigg| \ , 
\end{equation} 
where $\Delta M^\text{SM}_s$ is the SM prediction and $M^{Z^\prime}_{12}$ ($M_{12}^\text{SM}$) is the $Z^\prime$ (SM) contribution to the mixing amplitude. The modification due 
the $Z^\prime$ contribution to $M_{12}$ is given by
\begin{equation}
\frac{M_{12}^{Z^\prime}}{M_{12}^\text{SM}} = \frac{e^4}{2\pi^2} \frac{| C_9^\text{NP}|^2}{M_W^2 S_0(x_t)} \frac{m^2_{Z^\prime}}{(q_\ell g^\prime)^2} 
\end{equation}
where $S_0(x_t)$ us an Inami-Lim function \cite{Inami:1980fz} whose value is $\simeq 2.3$, with  $x_t = m_t/m_W$ being the ratio of the top quark mass to the $W$ boson mass. Bounds on generic new physics contributions to $\Delta M_s$  allow for $\Big| M_{12}^{Z^\prime}/M_{12}^\text{SM}\Big| \lesssim 0.12 $ \cite{Charles:2020dfl}, and will produce an upper bound on $m_{Z^\prime}$ as a function of $g^\prime$ once $C_9^\text{NP}$ is fixed. This bound is shown in Fig.~(\ref{fig:DD}) by the gray shaded region in the bottom-right corner of the plots labeled ``$B_s$ Meson Mixing''.

\subsection{Collider Constraints}

A powerful probe of the $Z^\prime$ parameter space is neutrino trident production -- the production of muon pairs when a muon neutrino scatters with the Coulomb field of a target nucleus. This most precise measurement of this process is from the CCFR \cite{CCFR:1991lpl} collaboration which finds good agreement with the SM \cite{Altmannshofer:2014cfa}
\begin{equation}
\sigma_\text{CCFR} / \sigma_\text{SM} = 0.82 \pm 0.28 \, .
\end{equation}
This result can set strong constraints on the parameter space of the $\LmuLtau$ model. Following the procedure in \cite{Altmannshofer:2014pba}, the region of parameter space excluded by neutrino trident production is depicted in Fig.~(\ref{fig:DD}) by the gray shaded region above the line labeled CCFR.

Additional constraints on the parameter space of this model can be obtained from measurements of $Z \to 4 \mu$ decays when the $Z^\prime$ is resonantly produced. A  search for this process was performed by the CMS collaboration \cite{CMS:2018yxg} and sets strong constraints on the $g^\prime$ coupling for 5 GeV $\gtrsim m_{Z^\prime} \gtrsim$ 70 GeV. These constraints are shown in Fig.~~(\ref{fig:DD}) by the gray curve labelled CMS. We see that CMS sets stronger constraints than CCFR in this region of parameter space.

Taking into account existing constraints from CCFR \cite{Altmannshofer:2014pba}, CMS \cite{CMS:2018yxg}, and $B_s$ meson mixing, and also requiring that $m_{Z^\prime} >$  10 GeV, the region favored by the $B$ anomalies is the central white region in Fig.~(\ref{fig:DD}) enclosed by the gray shaded regions above  $m_{Z^\prime} \simeq 50$ GeV.\footnote{Note, that there are additional constraints from $e^+e^- \to 4\mu$ searches from BaBar \cite{TheBABAR:2016rlg} but these apply outside of  region favored by the $B$ anomalies and we omit them from Fig.~(\ref{fig:DD}).}

\section{Dark Matter Phenomenology}\label{sec:DM}
In this section we discuss the relic density, direct detection, and indirect detection of DM in the model, motivated by the region of parameter where the $Z^\prime$ can explain the $B$ anomalies. In particular, we pay special attention to the role of the momentum dependence of the kinetic mixing in DM-nucleon scattering and the effect this has on DM direct detection constraints.

\subsection{The thermal dark matter target}
\begin{figure}[t]
\centering
\includegraphics[width=0.465\textwidth]{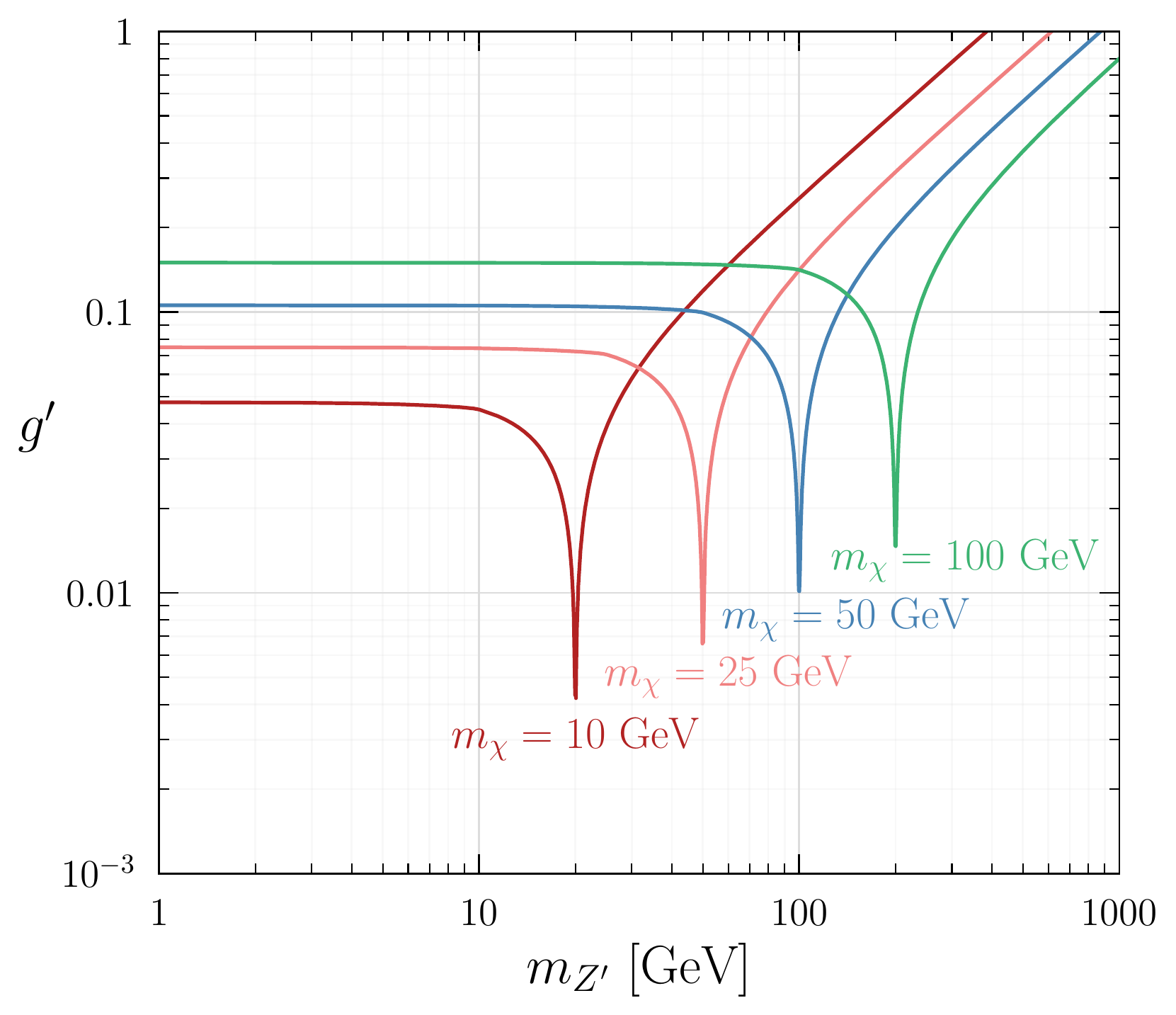}
\caption{Regions of the $g^\prime$ vs $m_{Z^\prime}$ parameter space where the DM obtains the observed relic abundance. The red, coral, blue, and green curves are for $m_\chi$ = 10, 25, 50, and 100 GeV, respectively.}\label{fig:relic}
\end{figure}

We consider a DM candidate $\chi$ whose relic abundance is obtained via freeze out. The relic abundance of DM today is determined by DM annihilation into $Z'$ bosons, or annihilation into SM leptons via off-shell $Z'$ exchange. The corresponding annihilation cross sections are
\begin{subequations}
\begin{align}
(\sigma v)_{\chi\bar\chi\to Z'Z'} &= \frac{g^{\prime4}}{16\pi m_\chi^2} \left( 1 - \frac{m_{Z'}^2}{m_\chi^2} \right)^{3/2} \left( 1 - \frac{m_{Z'}^2}{2m_\chi^2} \right)^{-2}  \label{eq:sigmaZZ}\ , \\
(\sigma v)_{\chi\bar\chi\to \nu_\ell \bar\nu_\ell} &= \frac{g^{\prime 4}  m_\chi^2}{2\pi [(4m_\chi^2 - m_{Z'}^2)^2 + m_{Z'}^2 \Gamma_{Z'}^2]}  \label{eq:sigmanu}\ , \\
(\sigma v)_{\chi\bar\chi\to \ell^+\ell^-} & = \frac{g^{\prime 4} m_\chi^2}{\pi [(4m_\chi^2 - m_{Z'}^2)^2 + m_{Z'}^2 \Gamma_{Z'}^2]}\nonumber \\
&~~\times \left( 1 + \frac{m_{\ell_a}^2}{2m_\chi^2} \right)  \sqrt{1 - \frac{m_{\ell_a}^2}{m_\chi^2}}  \label{eq:sigmaell} \ ,
\end{align}
\end{subequations}
where the flavor index $\ell=\mu$ or $\tau$. The $Z'$ decay width is~\cite{Kelly:2020pcy}
\begin{equation}
\Gamma_{Z'} = \frac{g^{\prime 2} m_{Z'}}{12\pi} \left[ 1 + \sum_{\alpha=\mu}^\tau (1+ 2 r_\alpha) (1-4r_\alpha)^{1/2} \Theta(1-4r_\alpha) \right] \ ,
\end{equation}
where $r_\alpha = m_\alpha^2/m_{Z'}^2$ and $\Theta$ is the Heaviside theta function. Given the annihilation cross sections, the DM relic abundance can be found by numerically solving the Boltzmann equation.

In Fig.~\ref{fig:relic} we show the region of parameter space where DM obtains the observed relic abundance of $\Omega h^2 \simeq 0.12$ \cite{Planck:2018vyg}. The red, coral, blue and green curves are for $m_\chi$ = 10, 25, 50, and 100 GeV, respectively. The behavior of these curves can be understood by analyzing the annihilation cross sections in different regimes. For $m_\chi \gg m_{Z^\prime}$ the annihilation cross section is proportional to $g^{\prime 4}/m^2_\chi$ and  $g^\prime$ is a constant for a fixed DM mass, as shown in the left part of the curves of Fig.~(\ref{fig:relic}). Then, we see the effect of annihilation near a resonance at $m_\chi \simeq 2 m_{Z^\prime}$, where the cross section is enhanced and a much smaller value of $g^\prime$ can produce the correct relic abundance. Finally, when $m_\chi \ll m_{Z^\prime}$ the annihilation cross section scales as $(g^\prime / m_{Z^\prime})^4$ and larger couplings are needed for larger $Z^\prime$ masses.

\subsection{Direct Detection}

Since DM does not couple directly to quarks, its scattering with nuclei of DM direct detection experiments occurs through the kinetic mixing between $Z^\prime$ and the SM photon, and depends on the choice of boundary condition for the total kinetic mixing. Here we consider DM direct detection constrains for the two different choices $\varepsilon_\text{UV} = 0 $ and $\varepsilon_\text{IR} = 0 $ discussed in Sec.~\ref{sec:model}.

The DM-proton scattering cross section is given by
\begin{equation}
\label{eq:directdetectioncrosssection}
\sigma_{\chi p} \approx \frac{ g^{\prime2} \varepsilon_{\rm tot}^2 e^2 \mu_{\chi p}^2}{\pi m_{Z'}^4} \ ,
\end{equation}
where $\mu_{\chi p} = m_\chi m_p/(m_\chi + m_p)$ is the reduced mass of DM-proton system,  and we've neglected the momentum transfer in the $Z'$ propagator. To account for the momentum transfer dependence, we consider the nucleus level DM scattering rate derived using the standard halo model

\begin{align}\label{eq:rate}
R  &= \frac{N_A M_A \rho_\odot g^{\prime 2} Z^2 e^2}{2\pi m_\chi} \nonumber\\
& \times \int_{E_{\rm th}}^{E_R^{\rm max}} d E_R \frac{\varepsilon^2_{\rm tot}(Q) F_{\rm Helm}^2(Q)}{(Q^2 + m_{Z'}^2)^2} \int_{v_{\rm min}(E_R)}^{v_{\rm esc}} d^3 v \frac{f(\vec{v})}{v} \ ,
\end{align}
where $N_A$ is the number of target nucleus, $M_A$ and $Ze$ are the mass and electric charge of the nucleus,
$F_{\rm Helm}$ is the Helm form factor \cite{Helm:1956zz,Jungman:1995df,Duda:2006uk,Hoferichter:2020osn}, and $Q \simeq \sqrt{2M_AE_R}$ for non-relativistic scattering.
The local DM mass density is $\rho_\odot=0.3\,{\rm GeV/cm^3}$ and the velocity distribution is Maxwellian
$f(\vec{v}) = C \exp(- |\vec{v} + \vec{v}_\odot|^2/v_0^2)\, \Theta(v_{\rm esc} - |\vec{v} + \vec{v}_\odot|)$,
where $\vec{v}$ is the DM velocity in the rest frame of the solar system, $v=|\vec{v}|$, $v_\odot=220\,{\rm km/s}$, $v_0=235\,{\rm km/s}$, $v_{\rm esc}=550\,{\rm km/s}$~\cite{Lin:2019uvt}, and $C$ is a normalization factor such that $\int d^3 v f(\vec{v})=1$.

The limits of integration over velocity and recoil energy in Eq.~(\ref{eq:rate}) are
\begin{equation}
v_{\rm min}(E_R)= \sqrt{\frac{M_A E_R}{2 \mu_{\chi A}^2}}  \ , \quad\quad E_R^{\rm max} = \frac{2 \mu_{\chi A}^2 (v_{\rm esc} + v_\odot)^2}{M_A} \ ,
\end{equation}
where $\mu_{\chi A} = m_\chi M_A/(m_\chi + M_A)$ is the reduced mass of DM-nucleus system, and $E_{\rm th}$ is the energy threshold of the detector under consideration.

To determine the direct detection bounds on the coupling $g^\prime$ we begin by first considering the $\varepsilon_\text{UV} = 0$. Because we are interested in the region of parameter space where $m_{Z^\prime} \gtrsim 1$ GeV, we can neglect the momentum transfer in propagator of the scattering cross section and directly translate the experimental upper limit on the DM-nucleon scattering cross section into limits on $g^\prime$  as a function of $m_{Z^\prime}$, for a for a fixed value of $m_\chi$, using Eq.~(\ref{eq:directdetectioncrosssection}).

\begin{figure*}[t]
\centering
\includegraphics[width=0.458\textwidth]{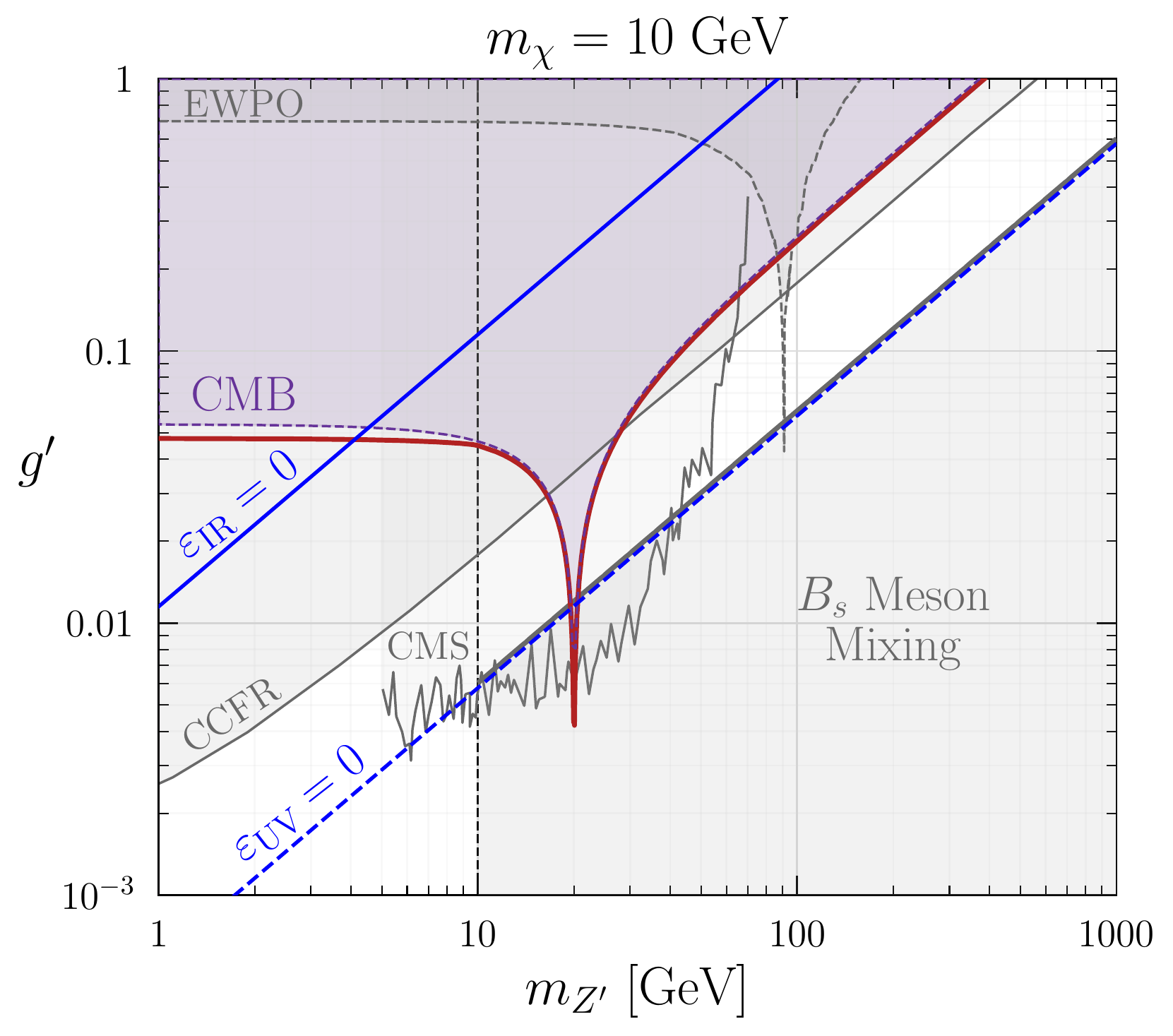}
\includegraphics[width=0.458\textwidth]{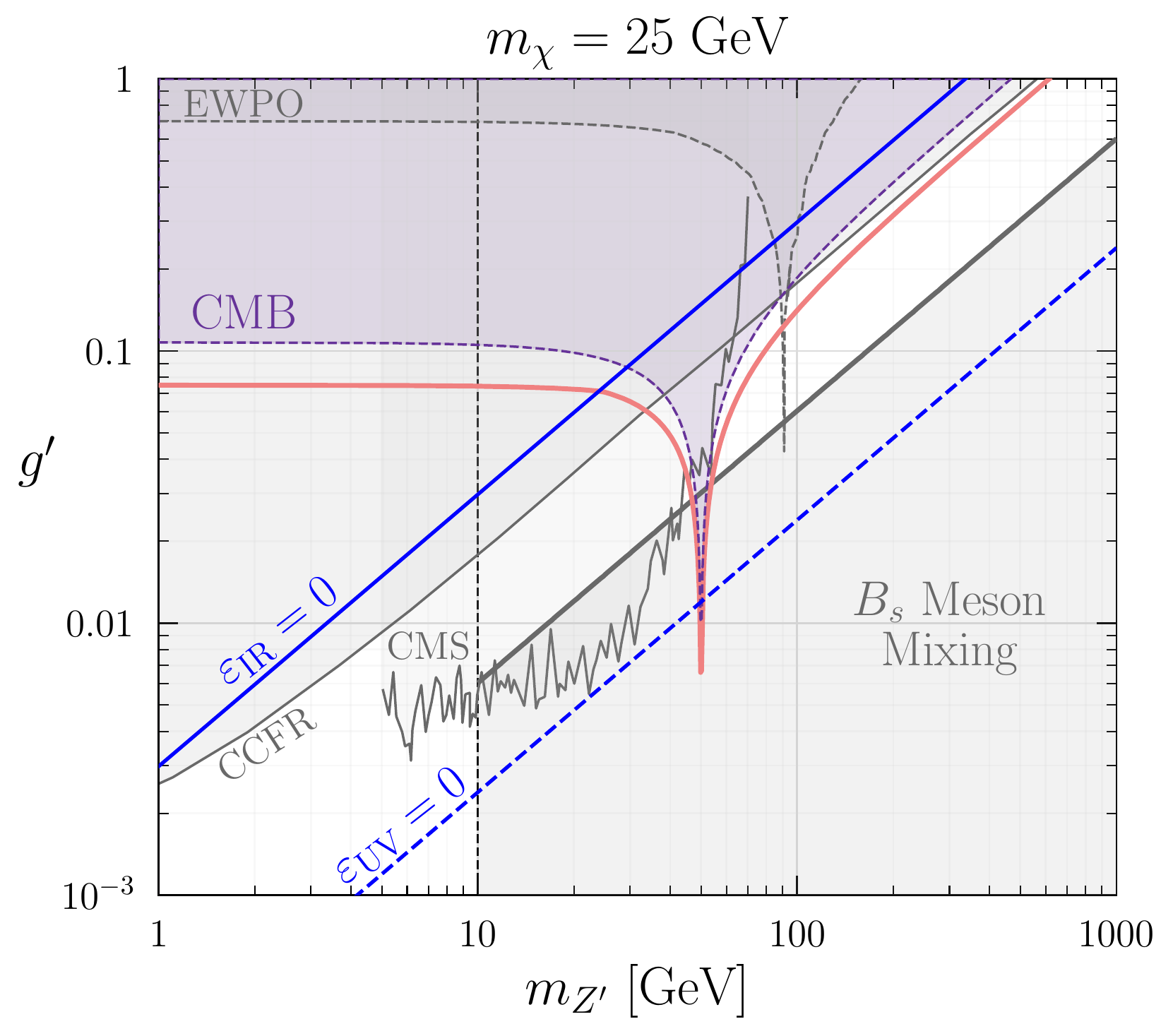}\\
\includegraphics[width=0.458\textwidth]{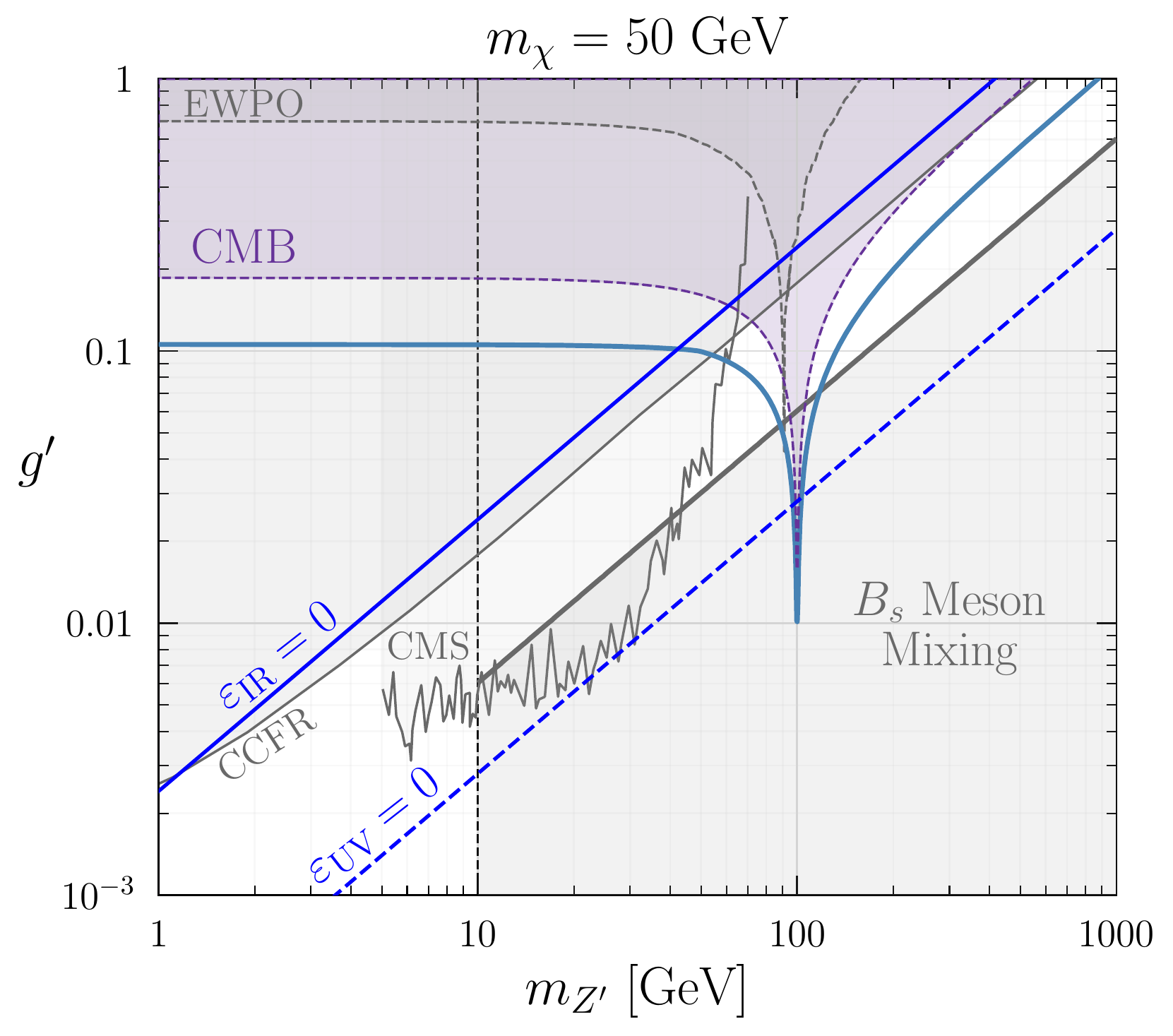}
\includegraphics[width=0.458\textwidth]{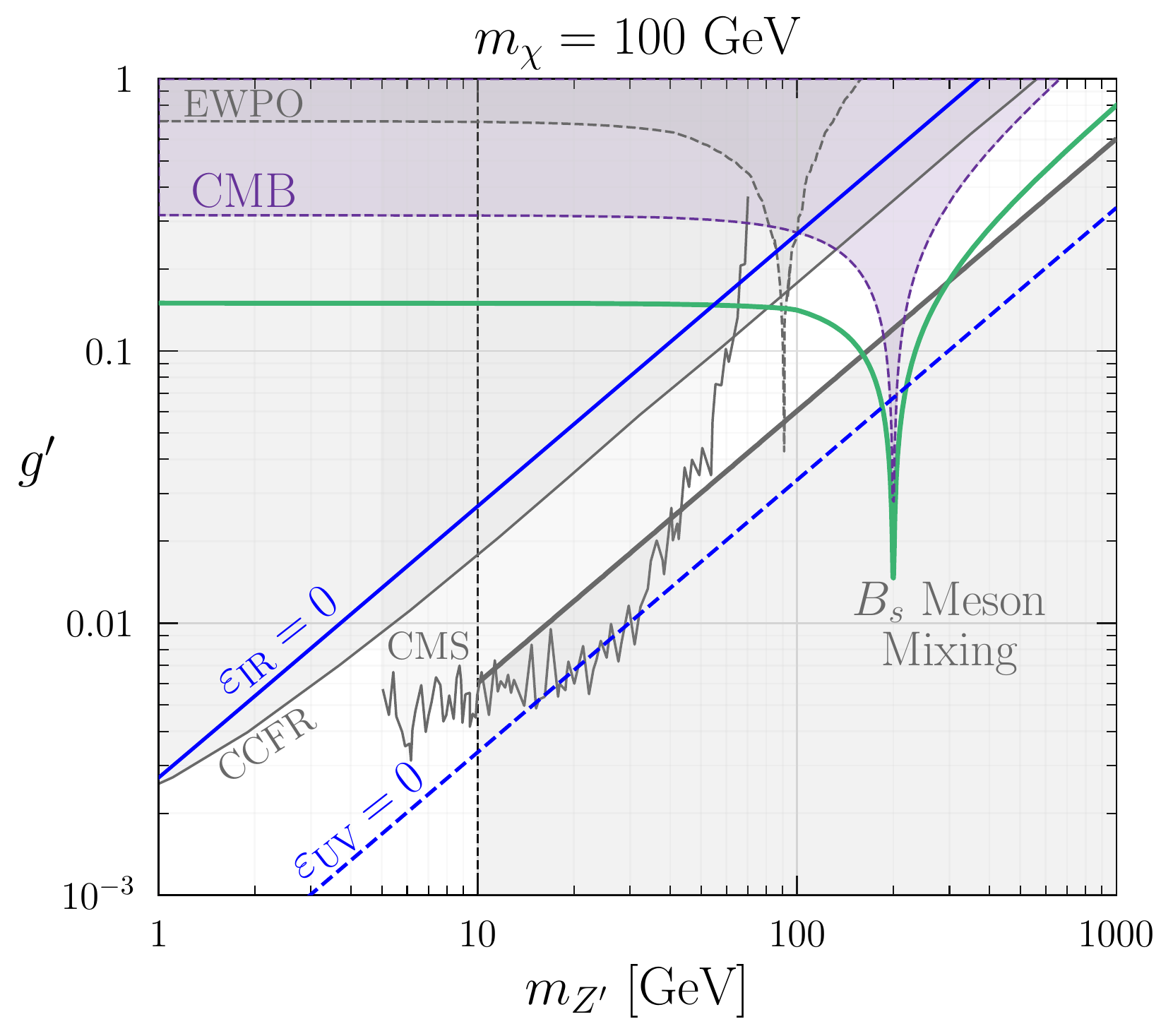}
\caption{Constraints of the $g^\prime$ vs $m_{Z^\prime}$ parameter space for $m_\chi=$ 10 GeV (top left),  25 GeV (top right), 50 GeV (bottom left), and 100 GeV (bottom right). The red, coral, dark blue, and green curves are the same as those of Fig.~(\ref{fig:relic}), where the DM obtains the correct relic abundance. The gray shaded regions enclosed by solid gray lines are existing constraints from neutrino trident production at CCFR \cite{Altmannshofer:2014pba}, a CMS search for $Z \to 4\mu$ \cite{CMS:2018yxg}, and from $B_s$ meson mixing. The dashed gray line is the upper bound from a shift in the $Z$ boson mass for  the $\varepsilon_\text{IR} = 0$ case. The purple shaded regions are constraints from energy injection into the CMB. The solid blue and dashed blue curves are DM direct detection constraints from LZ for $\varepsilon_\text{IR} = 0$ and $\varepsilon_\text{UV} = 0$, respectively.}\label{fig:DD}
\end{figure*}

Taking into account the strongest constraints from the LUX-ZEPELIN (LZ) experiment \cite{LUX-ZEPLIN:2022qhg}, the upper limit for $\varepsilon_\text{UV} = 0$ is depicted in Fig.~(\ref{fig:DD}) by the dashed  blue lines. We can see that direct detection constraints completely exclude the region of parameter space where the DM gets the relic abundance for all four DM masses that we consider, and that the model originally discussed in \cite{Altmannshofer:2016jzy} can no longer simultaneously address the $B$ anomalies and DM. 

\begin{figure}[t]
\centering
\includegraphics[width=0.465\textwidth]{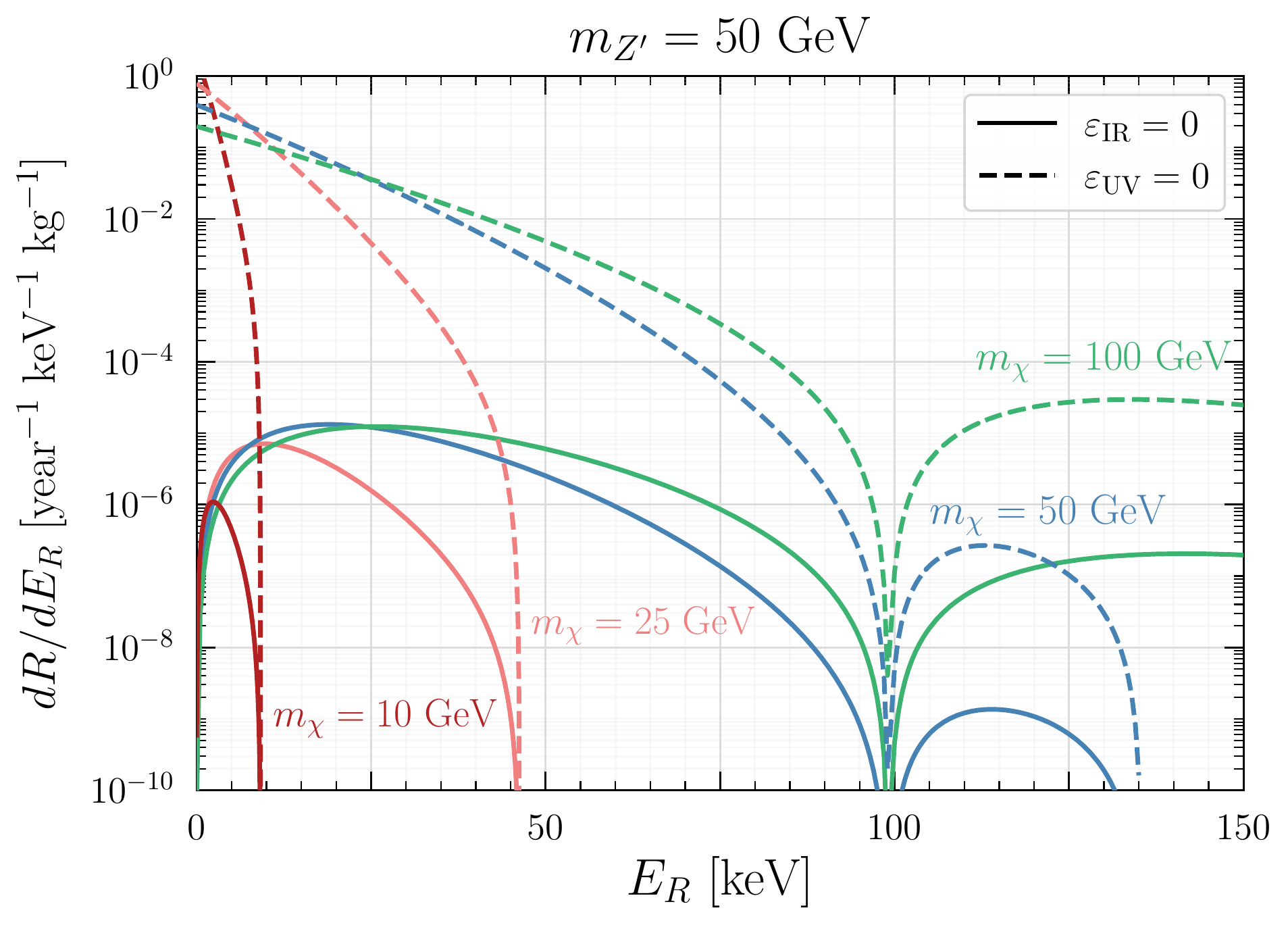}
\caption{Differential scattering rate of DM scattering with Xenon for the boundary condition $\varepsilon_\text{IR} = 0$ (sold curves) and  $\varepsilon_\text{UV} = 0$ (dashed curves), and fixed $m_{Z^\prime} =$50 GeV. The red, coral, blue, and green curves are for $m_\chi$ = 10, 25, 50, and 100 GeV, respectively. Note, the dip at $E_R \simeq 100$ keV is due to the behavior of the Helm form factor $F_\text{Helm}$ in the scattering rate and is independent of the model under consideration.}\label{fig:DiffRate}
\end{figure}

However, it can not be overemphasized that this conclusion holds only for the case where the kinetic mixing vanishes at high scales i.e. $\varepsilon_\text{UV} = 0$. This boundary condition of the kinetic mixing, as stated in Sec.~\ref{sec:model}, is not generic and the DM direct detection constraints can change significantly if the boundary condition is changed.

To this end, we now consider the case $\varepsilon_\text{IR} = 0$ where the kinetic mixing instead vanishes as low scales. To determine the upper bound on $g^\prime$  we need to rescale the bounds on the $\varepsilon_\text{UV}$ = 0 case by
\begin{equation}\label{eq:rescale}
\Bigg(\frac{R_{\varepsilon_\text{UV} = 0}}{R_{\varepsilon_\text{IR} = 0}} \Bigg)^{1/4}
\end{equation}
where $R$ is the scattering rate given by Eq.~(\ref{eq:rate}) evaluated for a particular boundary condition. Note, this ratio is independent of $m_{Z^\prime}$ once the DM mass is fixed. Furthermore, because the scattering rate for $\varepsilon_\text{IR} = 0$ is suppressed at low momentum transfer compared to $\varepsilon_\text{UV} = 0$, the ratio in Eq.~(\ref{eq:rescale}) is always greater than 1. This is illustrated in Fig.~(\ref{fig:DiffRate}) where we show the differential scattering rate $dR/dE_R$ for  $\varepsilon_\text{IR} = 0$ (solid curves) and $\varepsilon_\text{UV} = 0$ (dashed curves).  We fix $m_{Z^\prime} = 50$ GeV, and the different colored curves correspond to different DM masses. The suppression in the scattering rate can be seen clearly and implies that DM direct detection constrains will be weaker for $\varepsilon_\text{IR} = 0$ .

The results of this rescaling on DM direct detection constraints are depicted in Fig.~(\ref{fig:DD}) by the solid blue lines. We see that the DM direct detection constraints can be about an order of magnitude weaker for $\varepsilon_\text{IR} = 0$ compared to $\varepsilon_\text{UV} = 0$ case. Parameter space where DM obtains the correct relic abundance becomes viable for $m_\chi = $ 25, 50, and 100 GeV, and does not have to occur near the resonance $m_\chi \simeq 2 m_{Z^\prime}$. For $m_\chi = 10$ GeV we see that the relic abundance curve is excluded by a combination of existing constraints from CCFR \cite{Altmannshofer:2014pba} and CMS \cite{CMS:2018yxg}, but we keep it as an illustration of how the direct detection bounds vary when moving from $\varepsilon_\text{UV} = 0$ to $\varepsilon_\text{IR} = 0$.

To summarize, Fig.~(\ref{fig:DD}) shows that, by changing the boundary condition of the total kinetic mixing, we are able to open up parameter space where the $Z^\prime$ gauge boson can address the $B$ anomalies and, at the same time, serves as a mediator for DM that can obtain the correct relic abundance via thermal freeze out. Furthermore, the annihilation of DM in the early universe does not have to occur at the resonance, allowing a much wider parameter space for thermal DM.

\subsection{Indirect Detection}
The parameter space for $\LmuLtau$ charged DM can also be probed by indirect detection. In particular, for $m_\chi > m_{\mu,\tau}$ DM will annihilate to visible particles with a cross section given by Eq.~(\ref{eq:sigmaell}), and this can inject energy into the cosmic microwave background (CMB) and change the ionization history of the universe. Precise measurements of the CMB power spectrum can set strong constraints on the DM annihilation cross section \cite{Slatyer:2015jla,Planck:2018vyg,Kawasaki:2021etm}.

 These constraints are shown in Fig.~(\ref{fig:DD}) by the purple shaded regions. We see that for $m_\chi = 10$ GeV the CMB is probing almost all of the relic abundance curve, while for $m_\chi =$ 25, 50, and 100 GeV energy injection into the CMB is constraining only the region where DM annihilates on resonance $m_\chi \simeq 2 m_{Z^\prime}$. 

However, it is important to note that DM annihilation near a resonance is sensitive to the temperature at which the annihilation occurs. The annihilation cross section at the temperature of freeze-out can be different than when the CMB is formed \cite{Griest:1990kh,Gondolo:1990dk,Edsjo:1997bg,Ibe:2008ye,Guo:2009aj}. Therefore, the CMB constraints near the resonance should only be taken as an order of magnitude estimate. A more precise treatment is beyond the scope of this work, but will ultimately not have an effect on the CMB constrains outside of the resonance. 

\section{Electroweak Precision Constraints}\label{sec:EWPO}

In the previous section we considered the kinetic mixing between the $Z^\prime$ and SM photon since we are interested in DM direct detection, which is a low momentum process. Generally, however, the $Z^\prime$ will mix with the $Z$ boson at high scales and can shift the values of $m_Z$ and the $Z$ coupling to fermions away from their SM values. The shifts in the properties of the $Z$ boson can have an effect on predictions of electroweak (EW) precision observables, such as SM gauge boson masses and widths $m_{Z,W}$ and $\Gamma_{Z,W}$, the $Z$-peak hadronic cross section $\sigma^0_\text{had}$, $Z$ boson partial width ratios $R^0_{\ell, c, b}$, and forward-backward asymmetries $A^{f,0}_{FB}$.

Measurements of these observables from LEP and the LHC are in good agreement with SM predictions \cite{Baak:2014ora,Haller:2018nnx}, and can be used to constrain the parameter space of this model. A full treatment of the effect of $Z-Z^\prime$ mixing on EW precision observables is beyond the scope of this paper, but approximate constraints can be obtained by considering the effect on the predicted value of  $Z$ boson mass.

At high energies, where the $Z$ boson mass is precisely measured, the total kinetic mixing is approximately given by
\begin{equation}
\varepsilon_\text{tot}(Q) \simeq \varepsilon_0 + \frac{eg^\prime}{2\pi^2}\frac{m^2_\tau}{Q^2} \ .
\end{equation}
We can ignore the second term since $Q^2 \gg m^2_\tau$ at LEP and LHC energies. For the $\varepsilon_\text{UV} = 0$ case, we have $\varepsilon_0 = 0$ and there is no constraint from EW precision observables. On the other hand,  $\varepsilon_0$ is nonzero for the $\varepsilon_\text{IR} = 0$ case, and we can place constraints on $g^\prime$.

 To find the physical masses of the $Z$ and $Z^\prime$ gauge bosons, we need to diagonalize the kinetic terms in Eq.~(\ref{eq:Lag}) and the resulting neutral gauge boson mass-squared matrix. The physical masses are given by \cite{Curtin:2014cca}
\begin{align}\label{eq:masses}
m^2_{Z,Z^\prime} &= \frac{m^2_{Z,0}}{2} \Bigg(1 + \delta^2 + \eta^2\sin^2\theta_W \nonumber \\
&\pm \text{Sign}(1 - \delta^2) \sqrt{(1+\delta^2 +  \eta^2\sin^2\theta_W)^2 - 4 \delta^2}\Bigg),
\end{align}
where $m^2_{Z,0} = (g^2_1 + g^2_2)v^2/4$  is the $Z$ boson mass before mixing, with $g_1,g_2$ and $v\simeq 246$ GeV being the $U(1)_Y$ gauge coupling, $SU(2)_L$ gauge coupling, and the SM Higgs vacuum expectation value, respectively. The parameter $\delta$ is given by the relation $m^2_{Z^\prime,0} \equiv m^2_{Z,0} \times \delta^2$ where $m^2_{Z^\prime,0}$ is the $Z^\prime$ mass before mixing, and $\eta \equiv \varepsilon_0/\sqrt{1-\varepsilon_0^2}$. Finally, $\theta_W$ is the Weinberg angle.

To determine the constraints on the $\LmuLtau$ parameter space, we fix $m_{Z,0}$ to the PDG fit value of $m_Z = 91.1876 \pm 0.0021$ \cite{Workman:2022ynf}, that is we assume that the SM perfectly agrees with the experimental measurement. Allowing for a 2$\sigma$ shift in the measured value of the $Z$ boson mass, we obtain the upper limits shown in Fig.~(\ref{fig:DD}) by the dashed gray curves labeled ``EWPO".  We see that constraints from the shift in $m_Z$ are generally weaker than constraints from CCFR and CMS, except in the region where $m_{Z^\prime} \simeq m_Z$.

\section{Conclusions}\label{sec:conclusion}

In this work, we revisited the simultaneous explanation of the $B$ anomalies and DM in the $\LmuLtau$ extension of the SM presented in \cite{Altmannshofer:2016jzy} with a more general treatment of the kinetic mixing between the $Z^\prime$ and the photon. In general, the kinetic mixing depends on the momentum transfer of a particular process and different boundary conditions can lead to drastically different behavior of the total kinetic mixing.

Typically, it is assumed that the total kinetic mixing vanishes at high energy scales and we have seen in Fig.~(\ref{fig:DD}) that direct detection constraints, together with constraints from $B_s$ meson mixing, completely exclude the DM relic abundance curves, and a simultaneous explanation of $B$ anomalies and DM in this scenario is not possible.  However, we have seen that there is a freedom to choose the boundary condition of the total kinetic mixing, and the picture changes significantly if we instead require that the kinetic mixing vanishes at low scales. 

The scattering rate of low momentum transfer processes such as DM-nucleon scattering can be substantially reduced, and we saw that strong direct detection constraints on the parameter space of this model can be lifted, as shown in Fig.~(\ref{fig:DD}). We found that the parameter space that is favored by both the $B$ anomalies and DM becomes viable when considering this alternative boundary condition for the kinetic mixing.  These results emphasize the utility of treating the momentum dependence of the kinetic mixing in a way that is different than what is typically studied.


\acknowledgements
We thank Tim Hapitas for collaboration in the early stages of this work. We are grateful to Wolfgang Altmannshofer, Stefania Gori, and Yue Zhang for helpful discussions and valuable feedback on the manuscript. The work of DT is supported by the Arthur B. McDonald Canadian Astroparticle Physics Research Institute.



\bibliography{references}

\end{document}